\documentclass[runningheads]{llncs}
\usepackage{graphicx}
\usepackage{multirow}
\usepackage[font=small]{caption}
\begin{document}
\title{AD-Lite Net: A Lightweight and Concatenated CNN Model for Alzheimer's Detection from MRI Images \thanks{Accepted in ICPR'24}}
%
%
\author{Santanu Roy\inst{1,}\inst{2}\orcidID{0000-0001-6963-8019} \and
Archit Gupta \inst{2}\orcidID{0009-0002-6064-6383} \and
Shubhi Tiwari \inst{2}\orcidID{0009-0000-3023-5562} \and
Palak Sahu \inst{2}\orcidID{0009-0003-2570-4864}
}

\authorrunning{F. Author et al.}
%
\institute{$^{1}$Dept. of CSE, Pandit Deendayal Energy University (PDEU), Gandhinagar, india, \\$^{2}$Dept. of CSE, NIIT University, Rajasthan, India\\
\url{http://www.springer.com/gp/computer-science/lncs}\\
\email{santanuroy35@gmail.com}, gupta.archit2001@gmail.com, tshubhi2807@gmail.com, palak.sahu20@st.niituniversity.in}
\maketitle              
\begin{abstract}
Alzheimer's Disease (AD) is a non-curable progressive neurodegenerative disorder that affects the human brain, leading to a decline in memory, cognitive abilities, and eventually, the ability to carry out daily tasks. Manual diagnosis of Alzheimer's disease from MRI images is fraught with less sensitivity and it is a very tedious process for neurologists. Therefore, there is a need for an automatic Computer Assisted Diagnosis (CAD) system, which can detect AD at early stages with higher accuracy. Until now, numerous researchers have proposed several deep-learning models to detect AD efficiently from MRI datasets. However, most of their methods have deployed lots of pre-processing and image-processing techniques, which yields a lack of generalization in the model performance. In this research, we have proposed a novel AD-Lite Net model (trained from scratch), that could alleviate the aforementioned problem. The novelties we bring here in this research are, (I) We have proposed a very lightweight CNN model by incorporating Depth Wise Separable Convolutional (DWSC) layers and Global Average Pooling (GAP) layers. (II) We have leveraged a ``parallel concatenation block'' (pcb), in the proposed AD-Lite Net model. This pcb consists of a Transformation layer (Tx-layer), followed by two convolutional layers, which are thereby concatenated with the original base model. This Tx-layer converts the features into very distinct kind of features, which are imperative for the Alzheimer's disease. As a consequence, the proposed AD-Lite Net model with ``parallel concatenation'' converges faster and automatically mitigates the class imbalance problem from the MRI datasets in a very generalized way. For the validity of our proposed model, we have implemented it on three different MRI datasets. Furthermore, we have combined the ADNI and AD datasets and subsequently performed a 10-fold cross-validation experiment to verify the model's generalization ability. Extensive experimental results showed that our proposed model has outperformed all the existing CNN models, and one recent trend Vision Transformer (ViT) model by a significant margin.

\end{abstract}
\section{Introduction}
Alzheimer's Disease (AD) is a severe, and fatal neurodegenerative disease [1] that usually targets older individuals. The early signs of Alzheimer’s are forgetting recent events, language issues, having problems with reasoning and gradually it leads to loss of one’s ability to perform everyday tasks. AD occurs due to abnormal protein accumulation including beta amyloid plaques and tau tangles in the brain. These changes cause mental deterioration since nerve cells are lost gradually and the connections between brain cells and communication get disrupted. Alzheimer’s Disease International (ADI) has estimated that dementia affects more than 50 million people across the world [2], which is a term that refers to symptoms of brain impairment. AD is the leading cause of dementia and accounts for 60-80\% of cases. AD affects particular structures within the brain, the hippocampus [3], which is one of those first attacked by AD. Neural changes in the hippocampus’s anatomy can be identified by measuring its volume and form, as well as that of gray matter substance with highly advanced imaging techniques such as Computed Tomography (CT), Positron Emission Tomography (PET) and Magnetic Resonance Imaging (MRI). Out of all these image-acquiring techniques, MRI is the most frequently employed. Because it is noninvasive and easily available, moreover, it causes less radiation to the human body. Examining the alterations in the Cerebrospinal Fluid System (CFS) [4] aids in identifying the phase of AD. As this disease progresses, there is an enlargement in CFS region and reduction of the cerebral cortex and hippocampus. At present, there is no effective treatment available for Alzheimer's disease (AD), and the only way to prevent it is through early detection, as modern methods can only delay the course of progression. However, manually extracting and interpreting the features of Alzheimer's disease and furthermore, classifying them into different grades (from MRI images), is a very tedious and complex task for Neurologists. Hence, there is a need for an automatic CAD system, in order to detect AD efficiently from MRI images.

Various deep learning models have been widely employed recently by numerous researchers, in order to develop an automatic CAD system of AD detection from MRI images. Modupe Odusam et al. [5] proposed a pre-trained ResNet-18 which detects Alzheimer's disease from MRI images at an accuracy of around 98-99\%. However, they considered any two classes, thus, their classification problem (binary) was slightly lesser complex than the multi-class classification. Hadeer A. Helaly et al. [6] proposed a CNN model E2AD2C (trained from scratch) which is comprised of 3 convolutional layers, 2 Fully Connected (FC) layers, and 1 output layer. Their model architecture had less number of hyper-parameters (to train) and was inspired by the standard VGG-16 model. Nevertheless, they have deployed many pre-processing techniques, for example, over-sampling and under-sampling methods, data-augmentation, MRI filtering and normalization etc. prior to feeding the data into a classifier.  Shakarami et al. [7] proposed an AlexNet-SVM model in order to predict Alzheimer's disease from PET images. Their method encompasses four different steps. (I) First, 3D PET images are converted into 2D slices (or, images), (II) The pixels (in 2D slices) have values more than
150 are only passed through, otherwise avoiding all other pixels. (III) AlexNet-SVM model is utilized for the feature extraction part, and (IV) the final classification is done by the majority voting on slices. Although their method seems like a reasonable method, after converting 3D images into 2D slices, it may lose some important information, thus, it is not so feasible. K.G. Achilleos et al. [8] proposed a manual feature extraction method, in which they had computed Haralick texture [9] features for hippocampal atrophy which is the most vital part for predicting AD from MRI images. Moreover, they combined these hippocampal textures with their volume and subsequently, they applied all these features to a 10-fold cross-validation Decision Tree (for a 4-class classification task). Another potential direction of approaching this imbalanced MRI datasets is to deploy Weighted Categorical Cross Entropy (WCCE) [10] which assigns weights for every class which is inversely proportional to the number of images in that class. M. Masud et al. [11] have employed similar WCCE on top of a lightweight CNN model in order to resolve the issue of class imbalance from MRI datasets for AD detection. Besides that, many more related research works can be found in [12]-[15].

Another valid direction of this research could be leveraging new recent trends, that is, self-supervised models [16] or, attention-based models, in order to alleviate class imbalance problem from these MRI datasets. Numerous self-attention transformer models have been widely popular and proposed in the domain of NLP [17]. However, their equivalent model, i.e., Vision Transformer (ViT) [18], still is not an automatic choice for researchers in the domain of computer vision or image classification. The reason why still CNN outperforms ViT is that, ViT needs larger data in order to generalize well, however, in most of the medical image diagnoses, we have weakly supervised data or very limited imbalanced data. Moreover, unlike CNN model, ViT does not leverage a multi-scale hierarchical structure [16] which has a special significance for image classification. Therefore, numerous researchers [19],[20] come up with the idea of integrating both of the notions of ViT and CNN simultaneously. Recently, Byeongho Heo et al. [20] have proposed a Pooling-based ViT (PiT), which incorporates pooling layers in the ViT model. This leverages a multi-scale hierarchical architecture in the ViT, moreover, due to utilizing many pooling layers the number of hyper-parameters in PiT has been drastically reduced. Numerous researchers also tried to incorporate equivalent channel attention [21-23] named Squeeze Attention or, Swin Transformer [24] on top of CNN model, in order to improve the efficacy of AD detection from MRI images. Jiayi Zhu et al. [21] proposed a Sparse self-attention block in order to detect Alzheimer's disease at early stages, from MRI images. This ``Sparse self-attention block'' can reduce the elements (by $logN$) that can represent the overall features $N$. Therefore, overall, the computational complexity of their model (called BraInf) has been considerably reduced. Z. Liu et al. [22] have proposed a novel Multi-Scale Convolutional Network (MSC-Net) comprising four parallel concatenations of convolutional layers with varying dilation rates. Additionally, they have integrated an attention module ``SE-Net'' into their MSC-Net to enhance channel independence.  

We have observed that most of the aforementioned state-of-the-art models [5-14] struggle to generalize across different MRI datasets for Alzheimer's detection. These models particularly exhibit overfitting when dealing with imbalanced and small datasets. Researchers utilize image processing techniques as pre-processing methods [6,7] to augment datasets in order to improve the efficacy of the deep learning model. However, while these techniques may work well on a specific dataset, they do not ensure effective generalization across diverse datasets. Furthermore, several attention modules [21-24] proposed for AD detection could not directly address the issue of class imbalance. Therefore, in this research, we aim to develop a lightweight CNN model (trained from scratch), specifically designed for Alzheimer's detection, such that it can alleviate the class imbalance problem and generalize well across diverse MRI datasets. 

\subsection{MRI Images Dataset and Its Challenges}
For extensive experimentation, we have employed 3 MRI datasets which are readily available on Kaggle. The first dataset of Alzheimer's Disease [25] contains a total of 5000 images which are labeled further into 4 classes – Mild-Demented, Moderate-Demented, Non-Demented, and Very Mild-Demented. We call this dataset ``Alzheimer’s Detection (AD) dataset''. Here, Moderate-Demented is severely demented and is analogous to Alzheimer's Disease (AD). Whereas, Mild-Demented and very Mild-Demented are early stages of Alzheimer's Disease. The number of images in Mild-Demented, Moderate-Demented, Non-Demented, and Very Mild-Demented are 717, 52, 2560, and 1792 respectively. A Second dataset, named ``ADNI-Extracted-Axial’’, consists of 2D axial images extracted from the Nifti ADNI from ADNI website [26]. This ADNI dataset is the most authentic MRI dataset for Alzheimer’s disease, followed by numerous researchers. This ADNI contains 5000 images which are further divided into 3 classes - Alzheimer’s Disease (AD), Mild Cognitive Impaired (CI), and Common Normal (CN). A third dataset OASIS [26] of four classes, is also utilized in this research. The number of images in Mild Dementia, Moderate Dementia, Non- Dementia, and Very mild Dementia are 5002, 488, 67200 and 13725 respectively. Hence, this is a huge class imbalance problem and conventional CNN models' efficacy may suffer due to the lack of generalizing ability in the minor classes.

\subsection{Contributions}
The contributions of this paper are as follows:
\begin{enumerate}
        \item A very lightweight CNN model, AD-Lite Net, has been proposed as a base model for detecting Alzheimer's disease efficiently, from MRI images dataset. 
	\item A ``parallel concatenation block'' is incorporated on top of this base model in order to alleviate the class imbalance problem and to increase generalization ability of the model. In this ``parallel concatenation block'' (pcb), one Transformation layer (Tx-layer) is employed which enables the model to extract distinct and complementary features which were essential for Alzheimer's detection. 
 \item A mathematical analysis of the proposed model AD-Lite Net is presented in this research. In this analysis, one new \textit{lemma} has also been proposed.  
 
 \item For validity purpose, the proposed AD-Lite Net has been implemented on three different MRI image datasets. Moreover, we merged the ADNI and AD datasets and subsequently conducted a 10-fold cross-validation experiment to test the model's generalization ability.  
\end{enumerate}

\section{Methodology}
This methodology section can be further divided into two parts: (a) Alzheimer's Detection Lite Network (AD-Lite Net), (b) Mathematical Analysis of AD-Lite Net.

\subsection{Alzheimer's Detection-Lite Network (AD-Lite Net)}
The proposed AD-Lite Net model is explored in Fig.1.  The proposed model is comprised of main two parts: (I) Main backbone CNN model (which is a very lightweight model or base model), (II) One parallel concatenation block is leveraged into this backbone CNN model in order to increase the generalization ability of the model. Overall, in the proposed framework, a total of 7 convolutional layers and two Depth-wise Separable Convolutional (DWSC) layers [27] are employed, as shown in Fig.1. The number of filters deployed in the backbone model are 16, 32, 64, 96, and 128 from the $1^{st}$ to $5^{th}$ convolutional layer respectively. Every convolution layer has the same kernel size 3$\times$3 (except the $1^{st}$ one having kernel size 5$\times$5) with zero padding ``same''. ReLU activation function is employed in all the convolutional layers,
whereas, SoftMax activation function is incorporated in the output of the
CNN model. Each convolutional layer is followed by a Max-pooling layer, which down-samples the image size by half, because of using stride 2. Subsequently, a batch normalization layer is also incorporated after every Max-pooling layer or convolution layer, in the model. This batch normalization layer converts the scattered 2D tensor input (after convolution) into a normalized distribution having mean $0$ and standard deviation $1$. It ensures a smooth gradient flow throughout the network and hence, reduces the over-fitting problem, to a certain extent.  
\begin{figure*}[h]
		\centering
		\includegraphics[width=11.5cm,height=5.6cm]{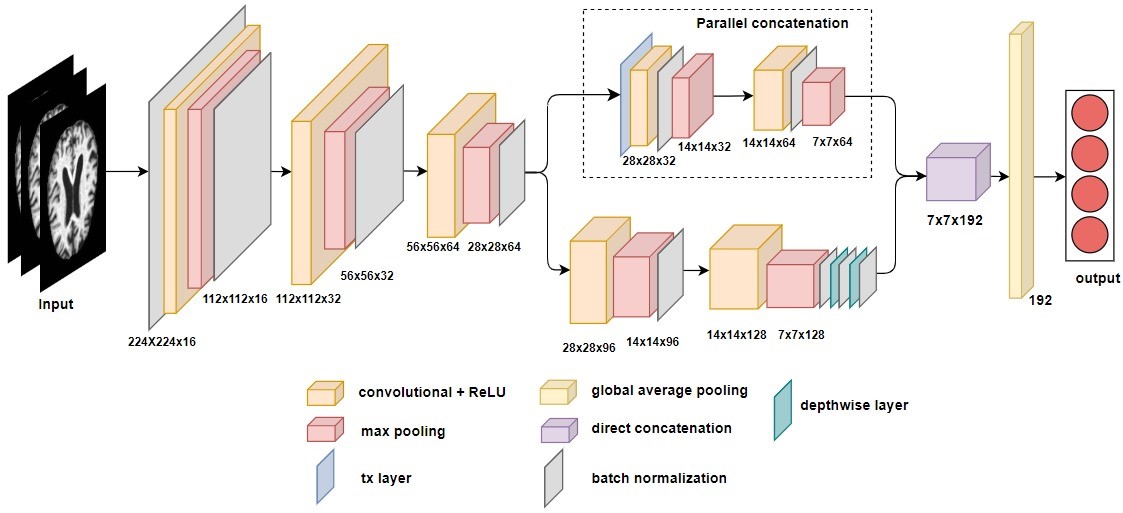}
		\caption{Block diagram of the proposed model AD-Lite Net}
	\end{figure*}

The ``parallel concatenation'' block starts from a transformation layer (or, Tx-layer) which converts the tensor output (coming from the $3^{rd}$ convolutional block) into a very different kind of image (i.e., negative image). This is further shown in Fig.2. This tx-layer is further followed by 2 convolutional layers and 2 Max-pooling layers. These two back-to-back convolutional layers have the number of filters 32 and 64 respectively. These numbers are chosen empirically, which is further explored in an ablation study in Supplementary material. This parallel concatenation block (pcb) can work like like an equivalent 'Attention block' in the CNN model, which is further exploited in the next subsection. Thereafter, these two parallel blocks are concatenated by a concatenation block which is followed by Global Average Pooling (GAP) Layer [28] and output layer, as shown in Fig.1. This is to clarify that DWSC layers and GAP layers (instead of flatten layer) have reduced the computational complexity of the AD-Lite Net considerably. Moreover, due to avoiding the entire dense layer part, the number of hyper-parameters of this AD-Lite Net is reduced to only 2.3 lakhs (approximately), hence, the proposed framework can work efficiently even on a very small and imbalanced dataset without being affected much by overfitting.

\subsection{Mathematical Analysis of AD Lite-Net}
A mathematical analysis of AD Lite-Net is presented in this section, in order to understand the credibility of the proposed research with much clarity. 

\vspace{0.1cm}
The convoluted tensor output (after any convolutional layer) in our proposed model, can be represented by 

\begin{equation}
		O_{i}(f)_{w\times w}={ReLU(({\sum_{j=1}^{p_i}{{O_j(z)}_{3\times 3}}*{I(f))}_{w\times w}}+b)}  
\end{equation}
Here in equation (1), $p_i$ is the number of filters in the current convolutional layer, $I(f)$ is the original image having size w$\times$w, ${O_j(z)}$ is the convolutional filter, having kernel size 3$\times$3 or 5$\times$5, and the same stride=1, with zero padding ``same". Thus, the size of the convoluted output will be also the same, i.e., $w\times w$, $b$ is the bias, `$*$' in equation (1) indicates convolution operation. 

The number of hyper-parameters $h_{c,i}$ in this $i^{th}$ convolutional layer can be computed by the following equation.

\begin{equation}
		h_{c,i}= (3^{2}. p_{i-1}+1).p_{i}
\end{equation} 
Here, in equation (2), $p_{i-1}$ is the number of filters in the previous layer, `$.$' indicates point-wise multiplication.

On the other hand, the number of hyper-parameters $h_{D,i}$ in this $i^{th}$ DWCS layer is represented in equation (3). Comparing equation (2) and (3), we can conclude that $h_{D,i}<<h_{c,i}$ if $p_{i}$ is higher, because DWSC  utilizes only one $3\times3$ convolution layer followed by $1\times1$ layers [27] that does point-wise multiplication. 

\begin{equation}
		h_{D,i}= (3^{2}. 1+1).p_{i-1}=10p_{i-1}
\end{equation}
We have employed 2 such DWSC layers at the last block (as shown in Fig.1) such that the number of hyper-parameters will not be raised significantly.

The Max-pooling with stride 2 (and pool size 2$\times$2), is a down-sampling operation [18] that would reduce the original image size to its half. After utilizing a total $n$ number of Max-pooling layers, the tensor output will be
\begin{equation}
		(Max^{n}(I(f))_{w\times w})_{2\times 2|2}= (O_n(f))_{(w/{2^n}. w/{2^n})}
\end{equation}

Here, in our proposed model, $n=5$. Thus, the spatial dimension of the output will be $224/2^{5}$ x $224/2^{5}= 7\times7$. The spectral dimension in this last block is $(64+128)=192$, shown in the Fig.1. This last layer is passed through the GAP layer, instead of flatten layer. This GAP layer [28] takes an average in the spatial dimension, thus, the number of neurons in this GAP is reduced to $192$ only. Whereas, the number of neurons in the flatten layer would be $7\times7\times 192$. Thus, the number of neurons has decreased considerably, after leveraging GAP in the proposed model. This will have a significant impact on the total number of hyper-parameters in the model. Hence, this can be concluded that the proposed CNN model is indeed a very lightweight model, and it has a very less number of hyper-parameters (2.3 lakhs only), as compared to other existing CNN models (trained from scratch). 

\begin{figure*}[h]
		\centering
		\includegraphics[width=11.6cm,height=5.6cm]{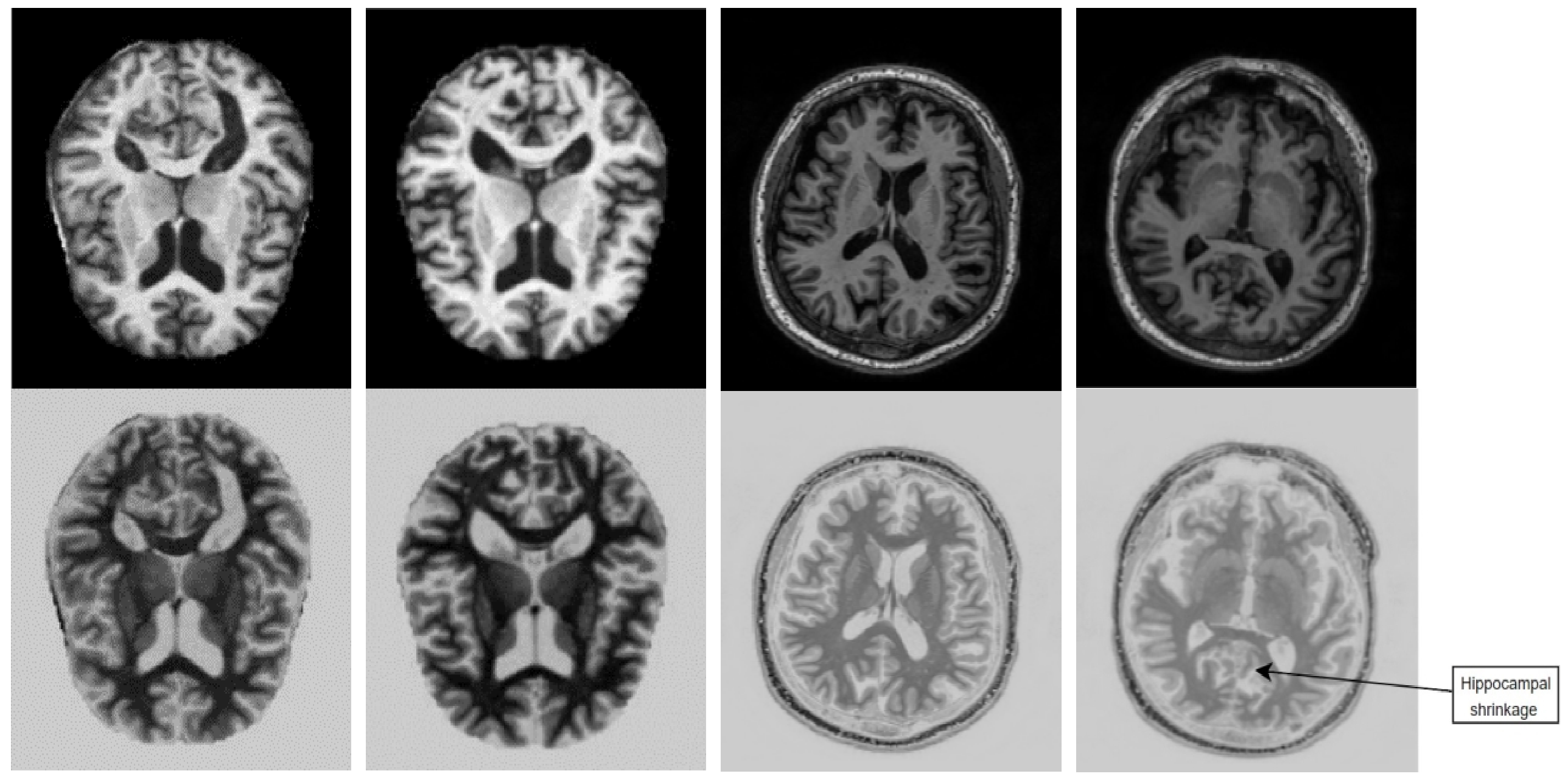}
		\caption{First row represents the original MRI images, 2nd row represents the images after passing it through Tx-layer}
	\end{figure*}
 
\vspace{0.1cm}
The parallel concatenation block (pcb) is one of the novelties of our research, shown in Fig.1. This pcb starts from a transformation layer (tx-layer), output of this tx-layer $I_o(f)$ is given in equation (5). This tx-layer is followed another 2 convolutional layers and two Max-Pooling layers, as shown in Fig.1.

\begin{equation}
		I_o(f)_{w x w}=m*(255-{I(f)}_{w\times w})
\end{equation}
where, $I_o(f)$ is the output of that transformation layer, $I(f)$ is the input tensor coming to the transformation layer, $m$ here is a real constant whose value is supposed to be $0<m<1$, empirically we have chosen the value of $m=0.8$ in this research. The purpose of this layer is to present the MRI images in such a format that it can highlight some hidden features which was not so prominent previously in the input tensor. In other words, it converts the original images into its negative version, such that it can extract additional essential features for Alzheimer's detection. We have further ensured that with medical hospital doctors. For instance, this is evident from Fig.2 (first two images) that the gray matter substance in hippocampus’s anatomy [3] of the original MRI image is more prominent after passing it through this Tx-layer. Similarly, in the last two images, in Fig.2 it has been highlighted that hippocampus shrinking [4] is more clear in the $2^{nd}$ row. Moreover, abnormal levels of beta-amyloid [4] and widespread deposits of this protein becomes more visible after passing the MRI images through this Tx-layer, according to the neurologists. These are significant features of AD that get more highlighted after utilizing the Tx-layer. 

\vspace{0.2cm}
The significance of Tx-layer in pcb, is explained in the following:
\begin{enumerate}
\item It can be observed from Fig.2 that the regions in the original image which were white, become more prominent and clear after passing through the Tx-layer. In contrast, areas in the transformed images that have changed to white (previously it was black in the original) become less prominent. Hence, it can be concluded that these two pairs of images (original and Tx-layer images) possess kind of complementary features. After consulting with neurologists, we came to know that this complementary features also carry some important information for AD detection. Therefore, incorporating both combinations of these features, enables the CNN model to learn more distinct and essential feature maps (for AD detection) than previous.

\item Moreover, it is evident from Fig.2 that the overall statistics in the original image and the processed image (i.e., after passing it through Tx-layer), differ significantly, thus, pcb may work like an efficient data augmenter inside the model. According to the research in [29], an efficient data-augmenter must generate synthetic images which have slightly different statistics compared to original images, otherwise, it induces overfitting in the model performance. 
\item Numerous researchers [21-23] proposed attention module in the form of parallel concatenation in their CNN framework. However, none of their techniques deployed transformation layers before, hence, there is a possibility that redundant features (or, very similar features) might have been extracted in those parallel concatenation blocks, leading to overfitting in the model performance. Our proposed framework first time introduced the concept of the Tx-layer (through pcb), which automatically transforms original feature maps into its complementary version. Thus, proposed pcb works like an efficient data (or, feature) augmenter inside the model, to the best of our knowledge. As a consequence, the proposed pcb block automatically increases the generalization ability of the model, thus, mitigating the class imbalance problem to a certain extent. 

\end{enumerate}

\vspace{0.1 cm}
We propose a new kind of \textit{lemma} of CNN model in this research, in a very generalized way which is as follows:

\vspace{0.1 cm}
\textit{Lemma1: If a CNN model, comprised of two parallel connections, extracts distinct features (in both such connections) that are essential for the final classification task, then that makes the model more stable than a series connection. Furthermore, extra distinct (or, complementary) features extracted in parallel concatenation, enable the network to generalize better for minor classes and thus, automatically alleviating the class imbalance problem efficiently.} 

\vspace{0.1cm}
 This is to clarify that, the idea of parallel concatenation is not exactly new. Previously Cornia Marcella et al. [30] pointed out one of the limitations of a Deep CNN model (having a large number of layers) that, the features that were extracted earlier at the beginning layers (of CNN), are mostly forgotten at the final decision of classification. Thus, many researchers suggested making a parallel concatenation to fuse those features from previous layers to the output layer. Later it becomes trends while numerous researchers [21-24] started employing attention module through parallel connection. In this research, we have furthermore extended that concept into a generalized concept that any CNN model, having those parallel concatenation layers, if extracting a bit distinct kinds of features, automatically resolves the class imbalance problem in a generalized way. For example, MobileNet-V2 [28], and Xception [27] models have already utilized similar kinds of parallel concatenation in their model architecture, therefore, they have decent performances on these imbalanced MRI datasets, despite having higher complexity of their architecture.

\section{Results and Analysis}
The results and analysis section can be further summarized into two, (a) Training specification, (b) Experimental results comparisons and analysis. 
\subsection{Training Specifications}
The training specifications of all of the models are given below:
	
\begin{enumerate}
\item The model was built using TensorFlow and Keras sequential API and the experiments were run on T4 GPU(Colab) environment as well as GPU P100(Kaggle). Colab environment provided a RAM of 25GB and Kaggle provided 100GB of RAM for the experiments.  
\item All the datasets were randomly split into 80-20\% ratio in a stratified way which is more feasible for class imbalance problem. This random splitting of train-test is the most authentic way of data splitting [31] so far for deep learning model. The train set was further partitioned into 80-20\% split (random) for creating the validation dataset. 
\item All the images in the entire dataset were resized to 224 x 224 prior to splitting the dataset. 
\item A Batch size of 64 was employed throughout all experiments to train all the CNN models. 
\item A learning rate (lr) of 0.00095 was chosen empirically, for Adams optimizer. 
\item For ``AD Dataset'', the model was trained for 18 epochs and moreover, an adaptive learning rate (alr) of 5\% decaying rate, is deployed after 8 epochs. 
\item For ``ADNI dataset'', we have not employed any alr, which means we train it for a fixed lr of 0.00095 for 15 epochs, because we have found ADNI (Axial) dataset is a very simple dataset and loss was converging much smoother way, without having any fluctuation.
\item For ``OASIS'' dataset, we employed a total of 7 epochs only, with alr (5\% decaying rate) employed after 4 epochs. 

\item This is to clarify that, we have not employed any early stopping criteria for model training, because we noticed that for a model (trained from scratch) early stopping often stops the training too earlier than expected. 

\item We have also implemented a pre-trained Pool-based Vision Transformer (PiT) model, on all three datasets. First, we have implemented it with the same training framework i.e., total 18 epochs with alr after 8 epochs. However, we observed that their model does not have the capability to learn very fast (in only 15 or 18 epochs). Thus, especially for PiT model, we also implemented the model for 50 epochs on all MRI datasets.
\end{enumerate}

\subsection{Experimental results comparisons and analysis}
We have implemented numerous pre-trained CNN models VGG-16, Xception, DenseNet-121, MobileNet etc. (which are 100\% fine-tuned from ImageNet dataset) on all three MRI datasets. Along with it, we have also implemented two existing CNN model, (I) 2D-M2IC (proposed by Helaly et al. [6]), and (II) MSC-Net (proposed by Liu, Z. et al. [22]) (trained-from-scratch) which were for AD detection. Furthermore, we have compared the efficacy of the proposed framework with a recent trend Pooling-based Vision Transformer (PiT) model [20]. Experimental results in Table 1, reveal that the proposed ``AD-Lite Net'' (trained from scratch) has consistently outperformed all the CNNs and PiT models by a substantial margin on all three MRI datasets. Furthermore, a comparison of the classification reports of the proposed AD-Lite Net model and AD-Lite without parallel concatenation, is presented in Table 2. This is to clarify that accuracy can not be counted on a specific class, it is always the overall accuracy of the model, thus, in Table 2 only one value of ``Accuracy'' is presented in one column. The results in Table 1 and Table 2 further strengthen and verify our proposed theory which was proposed in Section 2.2. Furthermore, the quality metrics along with their graphs, and confusion matrices of all these experiments (mentioned in table-1) are available in a Github link: 
\textbf{https://github.com/ArchitGupta16/Alzheimer-Detection/tree/main}.

An ablation study of the proposed AD-Lite model is also available in that link and this is further explored in a supplementary material.

\begin{table}[tb]
\begin{center}
\caption{Comparisons of several existing CNN models with the proposed framework (AD Lite-Net) on testing, for all three MRI datasets (Weighted Average)}
\label{tab:my-table}
\resizebox{0.98\columnwidth}{!}{
\begin{tabular}{|c|ccc|ccc|ccc|c|}
\hline
\multirow{2}{*}{\begin{tabular}[c]{@{}c@{}}Model/ \\ Methods\end{tabular}}                      & \multicolumn{3}{c|}{AD-Dataset}                                                                                                                                      & \multicolumn{3}{c|}{ADNI Dataset}                                                                                                                                    & \multicolumn{3}{c|}{OASIS Dataset}                                                                                                                                   & \multirow{2}{*}{\begin{tabular}[c]{@{}c@{}}No. of \\ param \\ (lakhs)\end{tabular}} \\ \cline{2-10}
                                                                                                & \multicolumn{1}{c|}{\begin{tabular}[c]{@{}c@{}}Accur-\\ acy\end{tabular}} & \multicolumn{1}{c|}{F1score}        & \begin{tabular}[c]{@{}c@{}}secs/\\ ep\end{tabular} & \multicolumn{1}{c|}{\begin{tabular}[c]{@{}c@{}}Accur-\\ acy\end{tabular}} & \multicolumn{1}{c|}{F1score}        & \begin{tabular}[c]{@{}c@{}}secs/\\ ep\end{tabular} & \multicolumn{1}{c|}{\begin{tabular}[c]{@{}c@{}}Accur-\\ acy\end{tabular}} & \multicolumn{1}{c|}{F1score}        & \begin{tabular}[c]{@{}c@{}}secs/\\ ep\end{tabular} &                                                                                     \\ \hline
\begin{tabular}[c]{@{}c@{}}DenseNet-121\\ (fine tuning)\end{tabular}                            & \multicolumn{1}{c|}{0.500}                                                & \multicolumn{1}{c|}{0.500}          & 48                                                 & \multicolumn{1}{c|}{0.739}                                                & \multicolumn{1}{c|}{0.739}          & 35                                                 & \multicolumn{1}{c|}{0.962}                                                & \multicolumn{1}{c|}{0.962}          & 76                                                 & 71.54                                                                               \\ \hline
\begin{tabular}[c]{@{}c@{}}VGG-16\\ (fine tuning)\end{tabular}                                  & \multicolumn{1}{c|}{0.648}                                                & \multicolumn{1}{c|}{0.627}          & 56                                                 & \multicolumn{1}{c|}{0.502}                                                & \multicolumn{1}{c|}{0.327}          & 44                                                 & \multicolumn{1}{c|}{0.251}                                                & \multicolumn{1}{c|}{0.167}          & 87                                                 & 147.17                                                                              \\ \hline
\begin{tabular}[c]{@{}c@{}}Xception\\ (fine tuning)\end{tabular}                                & \multicolumn{1}{c|}{0.892}                                                & \multicolumn{1}{c|}{0.891}          & 70                                                 & \multicolumn{1}{c|}{0.997}                                                & \multicolumn{1}{c|}{0.997}          & 54                                                 & \multicolumn{1}{c|}{0.951}                                                & \multicolumn{1}{c|}{0.951}          & 112                                                & 208.15                                                                              \\ \hline
\begin{tabular}[c]{@{}c@{}}MobileNet-V2\\ (fine tuning)\end{tabular}                            & \multicolumn{1}{c|}{0.938}                                                & \multicolumn{1}{c|}{0.938}          & 15                                                 & \multicolumn{1}{c|}{0.994}                                                & \multicolumn{1}{c|}{0.994}          & 12                                                 & \multicolumn{1}{c|}{0.973}                                                & \multicolumn{1}{c|}{0.973}          & 30                                                 & 32.11                                                                               \\ \hline
\begin{tabular}[c]{@{}c@{}}Pooling-based\\ ViT (PiT) {[}20{]}\end{tabular}                      & \multicolumn{1}{c|}{0.581}                                                & \multicolumn{1}{c|}{0.584}          & 20                                                 & \multicolumn{1}{c|}{0.621}                                                & \multicolumn{1}{c|}{0.618}          & 18                                                 & \multicolumn{1}{c|}{0.313}                                                & \multicolumn{1}{c|}{0.234}          & 30                                                 & 45.91                                                                               \\ \hline
\begin{tabular}[c]{@{}c@{}}Pooling-based\\ ViT (PiT) {[}20{]}\\ with 50 epochs\end{tabular}     & \multicolumn{1}{c|}{0.917}                                                & \multicolumn{1}{c|}{0.917}          & 20                                                 & \multicolumn{1}{c|}{0.925}                                                & \multicolumn{1}{c|}{0.926}          & 18                                                 & \multicolumn{1}{c|}{0.285}                                                & \multicolumn{1}{c|}{0.267}          & 30                                                 & 45.91                                                                               \\ \hline
\begin{tabular}[c]{@{}c@{}}2D-M2IC {[}6{]}\\ (train-from-scratch)\end{tabular}                  & \multicolumn{1}{c|}{0.882}                                                & \multicolumn{1}{c|}{0.881}          & \textbf{2}                                         & \multicolumn{1}{c|}{0.996}                                                & \multicolumn{1}{c|}{0.996}          & \textbf{1}                                         & \multicolumn{1}{c|}{0.937}                                                & \multicolumn{1}{c|}{0.937}          & 3                                                  & \textbf{8.19}                                                                       \\ \hline
\begin{tabular}[c]{@{}c@{}}MSC-Net+SE-Net {[}22{]}\\ (train-from-scratch)\end{tabular}       & \multicolumn{1}{c|}{0.893}                                                & \multicolumn{1}{c|}{0.901}          & 81                                                 & \multicolumn{1}{c|}{0.530}                                                & \multicolumn{1}{c|}{0.51}           & 64                                                 & \multicolumn{1}{c|}{0.877}                                                & \multicolumn{1}{c|}{0.877}          & 113                                                & 144.58                                                                              \\ \hline
\textbf{\begin{tabular}[c]{@{}c@{}}AD-Lite Net \\ (train-from-scratch)\\ proposed\end{tabular}} & \multicolumn{1}{c|}{\textbf{0.982}}                                       & \multicolumn{1}{c|}{\textbf{0.981}} & 5                                                  & \multicolumn{1}{c|}{\textbf{0.999}}                                       & \multicolumn{1}{c|}{\textbf{0.999}} & 4                                                  & \multicolumn{1}{c|}{\textbf{0.996}}                                       & \multicolumn{1}{c|}{\textbf{0.996}} & 12                                                 & 2.32                                                                                \\ \hline
\end{tabular}
}
\end{center}
\end{table}

\begin{table}[tb]
\begin{center}
\caption{Comparisons of Classification Reports of the proposed AD-Lite Net model with and without Parallel Concatenation, on the ``AD-Dataset''}
\label{tab:my-table}
\resizebox{0.9\columnwidth}{!}{
\begin{tabular}{|c|cccc|cccc|}
\hline
\multirow{2}{*}{Classes} & \multicolumn{4}{c|}{\begin{tabular}[c]{@{}c@{}}AD-Lite Net without \\ Parallel Concatenation\end{tabular}} & \multicolumn{4}{c|}{\begin{tabular}[c]{@{}c@{}}AD-Lite Net with \\ Parallel Concatenation\end{tabular}} \\ \cline{2-9} 
 & \multicolumn{1}{c|}{Precision} & \multicolumn{1}{c|}{Recall} & \multicolumn{1}{c|}{F1-Score} & \begin{tabular}[c]{@{}c@{}}Accuracy\\ \end{tabular} & \multicolumn{1}{c|}{Precision} & \multicolumn{1}{c|}{Recall} & \multicolumn{1}{c|}{F1-score} & \multicolumn{1}{l|}{\begin{tabular}[c]{@{}l@{}}Accuracy\\ \end{tabular}} \\ \hline
\begin{tabular}[c]{@{}c@{}}Very-Mild Demented\\ \end{tabular} & \multicolumn{1}{c|}{0.97} & \multicolumn{1}{c|}{0.94} & \multicolumn{1}{c|}{0.95} & \multirow{6}{*}{0.96} & \multicolumn{1}{c|}{1.00} & \multicolumn{1}{c|}{0.96} & \multicolumn{1}{c|}{0.98} & \multirow{6}{*}{\textbf{0.98}} \\ \cline{1-4} \cline{6-8}
\begin{tabular}[c]{@{}c@{}}Mild Demented\\ \end{tabular} & \multicolumn{1}{c|}{0.99} & \multicolumn{1}{c|}{0.94} & \multicolumn{1}{c|}{0.97} &  & \multicolumn{1}{c|}{0.99} & \multicolumn{1}{c|}{0.98} & \multicolumn{1}{c|}{0.99} &  \\ \cline{1-4} \cline{6-8}
\begin{tabular}[c]{@{}c@{}}Moderate Demented\\ \end{tabular} & \multicolumn{1}{c|}{1.00} & \multicolumn{1}{c|}{0.92} & \multicolumn{1}{c|}{0.96} &  & \multicolumn{1}{c|}{1.00} & \multicolumn{1}{c|}{1.00} & \multicolumn{1}{c|}{1.00} &  \\ \cline{1-4} \cline{6-8}
\begin{tabular}[c]{@{}c@{}}Non-Demented\\ \end{tabular} & \multicolumn{1}{c|}{0.95} & \multicolumn{1}{c|}{0.99} & \multicolumn{1}{c|}{0.97} &  & \multicolumn{1}{c|}{0.97} & \multicolumn{1}{c|}{1.00} & \multicolumn{1}{c|}{0.98} &  \\ \cline{1-4} \cline{6-8}
\begin{tabular}[c]{@{}c@{}}Macro-Average\\ \end{tabular} & \multicolumn{1}{c|}{0.98} & \multicolumn{1}{c|}{0.95} & \multicolumn{1}{c|}{0.96} &  & \multicolumn{1}{c|}{\textbf{0.99}} & \multicolumn{1}{c|}{\textbf{0.98}} & \multicolumn{1}{c|}{\textbf{0.99}} &  \\ \cline{1-4} \cline{6-8}
\begin{tabular}[c]{@{}c@{}}Weighted Average\\ \end{tabular} & \multicolumn{1}{c|}{0.96} & \multicolumn{1}{c|}{0.96} & \multicolumn{1}{c|}{0.96} &  & \multicolumn{1}{c|}{\textbf{0.98}} & \multicolumn{1}{c|}{\textbf{0.98}} & \multicolumn{1}{c|}{\textbf{0.98}} &  \\ \hline
\end{tabular}
}
\end{center}
\end{table}

\begin{figure*}[h]
		\centering
		\includegraphics[width=9.2cm,height=5.6cm]{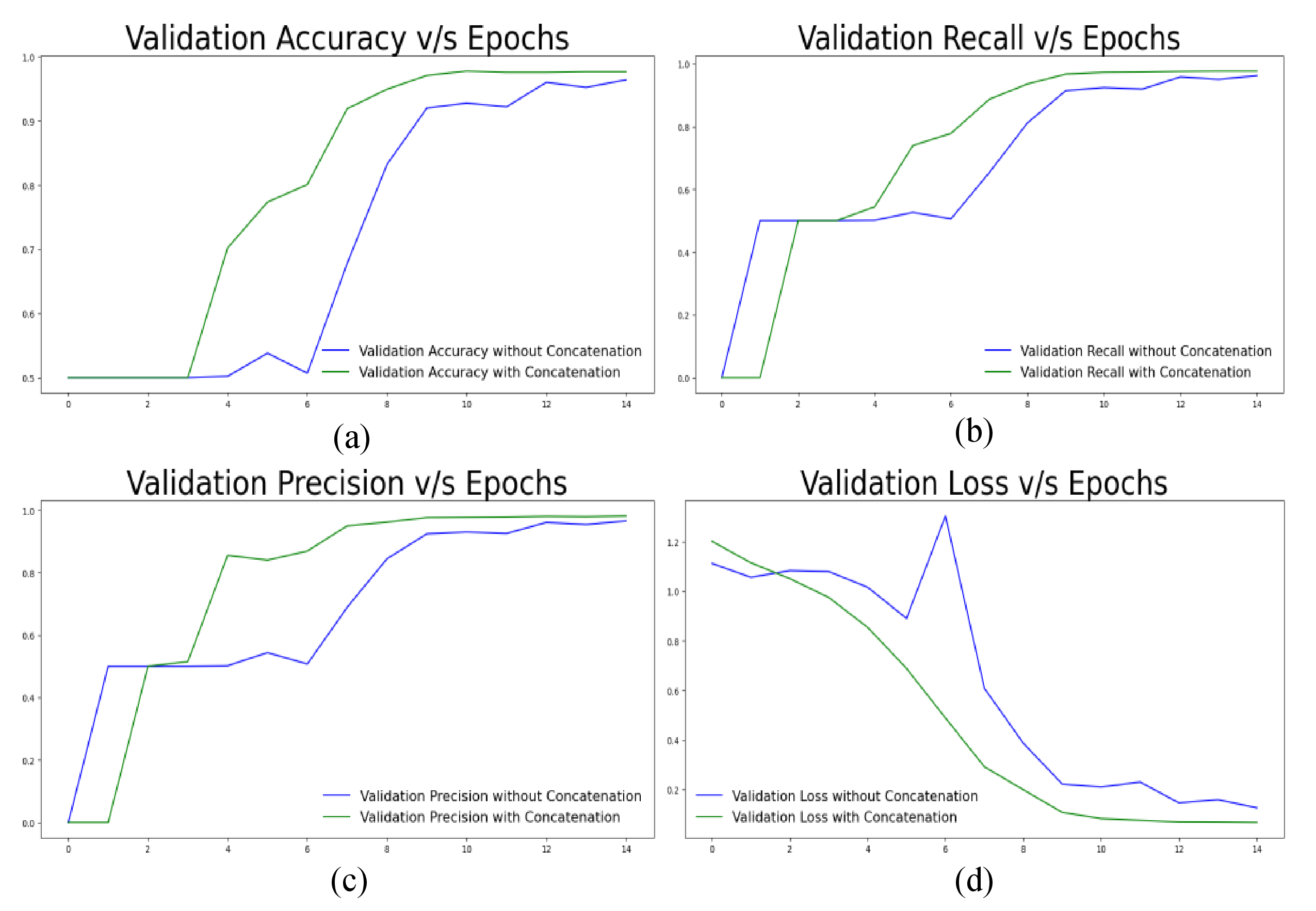}
		\caption{Validation graph comparison of proposed AD-Lite Net with vs AD-Lite Net without pcb, blue line indicates performance of AD-Lite without pcb and green line indicates performance of AD-Lite with pcb; (a) Accuracy vs epochs, (b) Recall vs epochs, (c) Precision vs epochs, (d) Loss vs Epochs}
	\end{figure*}

\vspace{0.1cm}
From Table-1 this is evident that the efficacy of the VGG-16 and Dense-Net are relatively lesser than that of other pre-trained CNN models. VGG-16 [32] usually does not deal well with the class imbalance problem, due to the lack of feature extraction in both spatial and spectral domains. Moreover, due to utilizing back-to-back convolutional layers (both in VGG-16 and DenseNet), the number of hyper-parameters in their model increased significantly, thus, over-fitting is inevitable in their model performances for small datasets. The most imbalanced dataset was the Oasis dataset, in which this is evident that VGG-16 suffers considerably to achieve higher accuracy and F1 score. Moreover, DenseNet-121 model suffers from very poor accuracy both in AD-Dataset and ADNI dataset. On the other hand, MobileNet-V2, Xception models have performed way better than VGG-16 and DenseNet-121, because of their lightweight framework. Xception is a modified version of Inception-V3 and the first time they incorporated Depth-Wise Separable Convolutional (DWSC) layers in their model, explored in Section 2.2. Whereas, MobileNet-V2 utilizes both DWSC layers and convolutional layers in its model, additionally, it leverages GAP layer instead of flatten layer. Due to utilizing these components in their model, both of these models avoid overfitting and as a consequence, they have decent performances throughout all these (small) MRI datasets.  

We have also implemented one of the recent trend models, Pooling based ViT (PiT) [20], on all three MRI datasets. Conventional ViT models can not be implemented on these small datasets, due to the complexity in their model architecture. Therefore, we have implemented PiT instead of ViT. From Table-1, this is apparent that the PiT model with 50 epochs, has achieved decent efficacy in both AD and ADNI datasets, however, their model has struggled to generalize in minor classes, for OASIS dataset. A recently proposed 2D CNN model (2D-M2IC) [6] is also implemented in this study, which is trained from scratch. The number of hyper-parameters in 2D-M2IC is considerably lesser (8.19 lakhs) than in other models. Table-1 shows that 2D-M2IC achieves good accuracy, and F1 score both in ADNI and OASIS datasets, however, it struggles to generalize the same in AD Dataset. Additionally, we have implemented a recently proposed model for AD detection, that is, ``MSC-Net,'' along with SE-Net attention block [22] on all three datasets. This model was trained from scratch with the same specification as the proposed model. Experimental results suggest that MSC-Net (with the SE-Net attention block) has achieved a commendable accuracy of 89.3\% and 87.7\% for AD and Oasis dataset respectively, nevertheless, it severely failed on ADNI dataset. Due to employing higher number of hyper-parameters (144.6 lakhs) it exhibited over-fitting for small dataset.

Overall, Table 1 reveals that some models performed occasionally well on particular datasets, however, most of them failed to generalize on all three MRI datasets. Only MobileNet-V2 [29], and the proposed AD-Lite Net model have obtained decent accuracy and F1 score more than 90\% consistently, over all three MRI datasets. Furthermore, this can be observed from Table 1 and Table 2 that the proposed AD-Lite Net has achieved the best accuracy, precision, recall, and F1 score (so far) on all three MRI datasets. This is also apparent from the graph in Fig.3 that the proposed ``AD-Lite Net'' has converged to higher accuracy and precision much faster after integrating the ``parallel concatenation block (pcb)''. This also reveals that by utilizing this pcb, the proposed framework generalizes much more effectively than previous and the validation graph becomes more stable. Furthermore, from Table 2, this is evident that the macro-averages of precision, recall, and F1-score have been boosted by 1-3\%, after leveraging pcb on the AD-Lite Net. This is a significant improvement, which justifies the necessity of incorporating ``pcb'' in the proposed framework. Hence, these experimental results support our proposed theory and \textit{Lemma1} which were proposed in Section 2.2.

We have also conducted a 10-fold cross validation experiment by combining two datasets. `AD dataset' and `ADNI dataset', which had dis-similar statistics. This merging is done after labelling the `Mild Demented' and `Very Mild Demented' classes in AD dataset into a single class Mild-Demented class. The idea was to blend diverse statistical images from these two datasets to create a challenging dataset. By this 10-fold cross-validation experiment, we effectively created the equivalent of 10 different datasets (we call them fold1-to-fold10 in Table-3), where each dataset has distinct testing set, having different statistics compared to the same of other 9 datasets. The results of this 10-fold cross-validation, with mean and standard deviation values, have been presented in Table-3 and also available in the aforementioned GitHub repository. These results demonstrate that the proposed ``AD-Lite Net'' is capable of achieving 98.3-99.7\% (Mean 99\%) accuracy consistently, in this challenging 10-fold cross-validation experiment as well. Furthermore, the standard deviation of accuracy, precision, recall and F1 score across these 10 folds is significantly low, that is 0.4\% only. This also indicates that the performance of the proposed model has been remarkably stable and it indeed resolved the class imbalance issue in a very generalized way. Hence, this experiment validates the generalization capability of the proposed model in a highly efficient way. 
\begin{table}[tb]
\begin{center}
\caption{Testing results for 10-fold cross validation on merged dataset}
\label{tab:my-table}
\resizebox{0.6\columnwidth}{!}{
\begin{tabular}{|c|c|c|c|c|c|c|}
\hline
\multirow{2}{*}{folds}                                  & \multirow{2}{*}{Accuracy}                              & \multicolumn{1}{l|}{\multirow{2}{*}{Precision}}        & \multicolumn{1}{l|}{\multirow{2}{*}{Recall}}           & \multirow{2}{*}{F1score}                               & \multicolumn{1}{l|}{\multirow{2}{*}{AUC}}               & \multicolumn{1}{l|}{\multirow{2}{*}{\begin{tabular}[c]{@{}l@{}}secs/\\ ep\end{tabular}}} \\
                                                        &                                                        & \multicolumn{1}{l|}{}                                  & \multicolumn{1}{l|}{}                                  &                                                        & \multicolumn{1}{l|}{}                                   & \multicolumn{1}{l|}{}                                                                    \\ \hline
fold1                                                   & 0.995                                                  & 0.995                                                  & 0.994                                                  & 0.994                                                  & 0.999                                                   & 7                                                                                        \\ \hline
fold2                                                   & 0.984                                                  & 0.984                                                  & 0.984                                                  & 0.984                                                  & 0.998                                                   & 6                                                                                        \\ \hline
fold3                                                   & 0.992                                                  & 0.992                                                  & 0.992                                                  & 0.992                                                  & 0.999                                                   & 7                                                                                        \\ \hline
fold4                                                   & 0.991                                                  & 0.991                                                  & 0.991                                                  & 0.991                                                  & 0.999                                                   & 6                                                                                        \\ \hline
fold5                                                   & 0.983                                                  & 0.983                                                  & 0.983                                                  & 0.983                                                  & 0.999                                                   & 6                                                                                        \\ \hline
fold6                                                   & 0.992                                                  & 0.992                                                  & 0.992                                                  & 0.992                                                  & 0.999                                                   & 6                                                                                        \\ \hline
fold7                                                   & 0.994                                                  & 0.994                                                  & 0.994                                                  & 0.994                                                  & 0.999                                                   & 6                                                                                        \\ \hline
fold8                                                   & 0.988                                                  & 0.988                                                  & 0.988                                                  & 0.988                                                  & 0.999                                                   & 7                                                                                        \\ \hline
fold9                                                   & 0.997                                                  & 0.997                                                  & 0.997                                                  & 0.997                                                  & 0.999                                                   & 6                                                                                        \\ \hline
fold10                                                  & 0.985                                                  & 0.985                                                  & 0.985                                                  & 0.985                                                  & 0.997                                                   & 7                                                                                        \\ \hline
\begin{tabular}[c]{@{}c@{}}Mean$\pm$\\ Std dev\end{tabular} & \begin{tabular}[c]{@{}c@{}}0.990$\pm$\\ 0.004\end{tabular} & \begin{tabular}[c]{@{}c@{}}0.990$\pm$\\ 0.004\end{tabular} & \begin{tabular}[c]{@{}c@{}}0.990$\pm$\\ 0.004\end{tabular} & \begin{tabular}[c]{@{}c@{}}0.990$\pm$\\ 0.004\end{tabular} & \begin{tabular}[c]{@{}c@{}}0.999$\pm$\\ 0.0008\end{tabular} & 6.4                                                                                      \\ \hline
\end{tabular}
}
\end{center}
\end{table}

\section{Conclusion and Future Work}
One lightweight and concatenated CNN model (train from scratch) was proposed for automatic Alzheimer's detection from MRI images. ``Parallel concatenation block'', incorporated into the base model, leveraged a novel Tx-layer which extracted unique salient features for Alzheimer's disease, thus, automatically mitigating the class imbalance problem in a generalized way. Experimental results on three different MRI datasets showed that there was a lack of generalization of all the existing and pre-trained CNN models. The AD-Lite Net model with concatenation block, not only generalized well for all three MRI datasets, but also, achieved the best accuracy, precision, recall, F1 score for all three datasets. Furthermore, the proposed framework outperformed one recent trends model, Pooling-based Vision Transformer (PiT), by a significant margin. Hence, this can be concluded that the proposed AD-Lite Net successfully alleviated all the challenges for AD detection from MRI datasets, and this proposed framework can perform well uniformly for any MRI dataset. A 10-fold cross-validation experiment also demonstrated the strong generalization capability of the proposed ``AD-Lite Net''.

This is to clarify that, until now, we worked with MRI datasets that did not include subject-specific images. Moving forward, our goal is to extend this project to predicting Alzheimer's disease at different subjects instantly which will be a more challenging and valid direction from the perspective of medical experts. In order to deal with more practical (noisy) data taken from a hospital, we are also planning to incorporate one extra attention module in our model.
%
%
%
%

\end{document}